\documentstyle{article}
\begin{document}

\def\dfrac#1#2{\frac{\displaystyle#1}{\displaystyle#2}}
\def\ov{\overline}
\def\tg{{\rm tg}}
\def\p{\partial}
\def\~#1{\stackrel{\sim}{#1}}

\begin{center}
\Large {\bf 
Binary Nonlinearization of AKNS Spectral Problem}\\
{\bf under Higher-Order
Symmetry Constraints}\\ 
\quad \\
\normalsize  {\large Yishen Li}\footnote{Email: ysli@nsc.ustc.edc.cn}\\
{\sl Department of Mathematics, City University of Hong Kong,
Kowloon, Hong Kong\\
Department of Mathematics and Center of Nonlinear Science,
\\ University of Science and Technology of China, Hefei 230026, China}
\vspace{1mm}\\
{\large Wen-Xiu Ma}\footnote{Email: mawx@math.cityu.edu.hk}\\
{\sl Department of Mathematics, City University of Hong Kong, Kowloon, 
Hong Kong}
\end{center}
\normalsize

\begin{center}
{\bf Abstract}\\
\vskip 0.2cm
\parbox{140mm}{\small Binary nonlinearization of AKNS spectral
problem is extended to the cases of higher-order symmetry constraints. 
The Hamiltonian structures, Lax representations, $r$-matrices and integrals
of motion in involution are explicitly proposed 
for the resulting constrained systems
in the cases of the first four orders.
The obtained integrals of motion 
are proved to be functionally independent and thus the constrained systems
are completely integrable in the Liouville sense.
}
\end{center}
\vskip 0.5cm

\section{Introduction}

Symmetry constraints have been applied to constructing finite-dimensional 
integrable systems from Lax pairs of soliton equations.
It has aroused an increasing interest 
among scientific researchers in the field of soliton theory
in the past several years. 
A large class of finite-dimensional integrable systems 
had been obtained indeed from soliton 
hierarchies by imposing
symmetry constraints$^{\cite{Cao-CQJM1988}-\cite{GengM-NCA1995}}$. 
There is also a $k$-constraint KP theory which uses symmetry constraints
to extend the standard KP 
theory$^{\cite{KonopelchenkoSS-PLA1991}-\cite{OevelS-CMP1993}}$.   
Nonlinearization method$^{\cite{Cao-CQJM1988}}$ utilizes 
a kind of special symmetry constraints for $1+1$ dimensional soliton
equations.
In refs.$^{\cite{MaS-PLA1994}-\cite{MaFO-PA1996}}$, one of the authors
suggested
binary nonlinearization method to get finite-dimensional integrable
system from Lax pairs of 1+1 dimensional integrable systems. 
This method has been successfully applied to a few of famous soliton 
hierarchies and it has to be adopted in making nonlinearization
when Lax pairs have odd-order matrix 
representations$^{\cite{MaFO-PA1996}}$.

However, what has been usually considered in binary 
nonlinearization is the Bargmann symmetry constraints
(explicit constraints). In this paper, we would like to 
extend binary nonlinearization method to the cases of higher-order symmetry
constraints (implicit constraints). 
The paper is organized as follows.
In Section 2, the Hamiltonian
 structures are constructed for the resulting constrained systems.
Main analysis is to describe how to introduce new dependent variables
to put the constrained systems to be Hamiltonian.  
In Section 3, the same $r$-matrices 
as those in mono-nonlinearization
are proved to be valid for 
the constrained
systems in binary nonlinearization after presenting Lax pairs.
In Section 4, the obtained integrals of motion 
are shown to be in involution in pairs and functionally independent.
It follows that
all those finite-dimensional dynamical systems are Liouville integrable.
A conclusion and some remarks are given in the final section. 

\section{The Hamiltonian structures}

For the sake of simplicity, 
we assume that the potentials 
$v,w$ and all their derivatives with respect to $x$ 
tend to zero when $|x|\rightarrow \infty$.

Following the notation in the reference$^{\cite{MaS-PLA1994}}$,
 we consider the spectral problem
$$
\left\{ \begin{array}{l}
\phi_{1x}=-\lambda\phi_1+v\phi_2,\vspace{2mm}\\
\phi_{2x}=w\phi_1+\lambda\phi_2,\end{array}\right.\ 
\textrm{i.e.}\  
\phi_x=U\phi,\  
\phi= \left(\begin{array}{c}\phi_1\vspace{2mm}\\ \phi_2\end{array}\right)
,\  U=U(u,\lambda)=\left(\begin{array}{cc}
-\lambda&v\vspace{2mm}\\
w&\lambda
\end{array}\right),\ 
u=\left(\begin{array}{c}
v\vspace{2mm} \\w
\end{array}\right)
\eqno{(2.1)}$$
and its adjoint spectral problem
$$
\left\{\begin{array}{l}
\psi_{1x}=\lambda\psi_1-w\psi_2,\vspace{2mm}\\
\psi_{2x}=-v\psi_1-\lambda\psi_2,\end{array}\right.\quad
\textrm{i.e.}\quad  
\psi_x=-U^T\psi,\quad \psi=
\left(\begin{array}{c}\psi_1 \vspace{2mm}\\ \psi_2\end{array}\right)
 \eqno{(2.2)}$$
for the AKNS soliton hierarchy.
Assume that $V_x=[U,V]$ has a formal solution
$$V= \left(\begin{array}{cc} a& b\\c&-a\end{array}\right)=\sum_{i=0}^\infty
\left(\begin{array}{cc} a_i& b_i\\c_i&-a_i\end{array}\right) 
\lambda ^{-i}, 
$$
where $a_i,\,b_i,\,c_i,\,i\ge 0,$ are determined uniquely by 
$$ a_{ix}=q c_i-r b_i,\  
 b_{ix}=-2b_{i+1}-2q a_i,\ 
c_{ix}=2c_{i+1}+2r a_i, \ i\ge0,$$
and 
$$ a_0=-1,\, b_0=c_0=0;\  a_i|_{u=0}=
b_i|_{u=0}=
c_i|_{u=0}=0,\ i\ge 1.$$
Upon making the Bargmann constraint, as in ref.$^{\cite{MaS-PLA1994}}$,
$$
\left(\begin{array}{c}
b_1\\
c_1\end{array}\right)
=\left(\begin{array}{c}
v\\
w\end{array}\right)=
\left(\begin{array}{c}
<\psi_2,\phi_1>\\
<\psi_1,\phi_2>
\end{array}\right)=
\left(\begin{array}{c}
\sum^N_{i=1}\psi_2(x,\lambda_i)\phi_1(x,\lambda_i)\\
\sum^N_{i=1}\psi_1(x,\lambda_i)\phi_2(x,\lambda_i)\\
\end{array}\right),
\eqno{(2.3)}$$
where we use $\phi_j,\,\psi_j,\,j=1,2,$ to denote the following 
vector valued functions 
$$\phi_j=(\phi_j(x,\lambda _1),..., \phi_j(x,\lambda _N))^T,\quad 
\psi_j=(\psi_j(x,\lambda _1),..., \psi_j(x,\lambda _N))^T,\quad j=1,2
 ,$$
we obtain from (2.1) and (2.2) the following Hamiltonian system
$$
P_x=-\frac {\partial H}{\partial Q},\quad Q_x=\frac {\partial H}{\partial P}
\eqno{(2.4)}$$
with the Hamiltonian 
$$H=<A\psi_2,\phi_2>-<A\psi_1,\phi_1>+<\psi_2,\phi_1><\psi_1,\phi_2>,\quad
A=diag(\lambda_1,...,\lambda_N),\ \lambda_i\not=\lambda_j\ (i\not=j),
\eqno{(2.5)}$$
where
$$\left.\begin{array}{l}
Q=(\phi_{1}(x,\lambda_1),...,\phi_{1}(x,\lambda_N),\phi_{2}(x,\lambda_1),...,
\phi_{2}(x,\lambda_N))^T,\\
P=(\psi_{1}(x,\lambda_1),...,\psi_{1}(x,\lambda_N),\psi_{2}(x,\lambda_1),...,
\psi_{2}(x,\lambda_N))^T.
\end{array} \right.$$

We will focus on the higher-order symmetry constraint
$$
\left(\begin{array}{c}
c_n\\
b_n\end{array}\right):=\frac {\delta H_{n-1}}{\delta u}
=\beta _0 \left(\begin{array}{c}
<\psi_1,\phi_2>\\<\psi_2,\phi_1>
\end{array}\right), \ H_{n-1}=\frac 2 {n}a_{n+1},
\eqno{(2.6)}$$
where $\beta_0$ is a constant to be determined and 
$n$ is a positive integer.
If $n$ is odd,
we can introduce the Jacobi-Ostrogradsky coordinates
to put the constrained system 
consisting 
of (2.1) and (2.2) with all $\lambda =\lambda _i$ and (2.6)
to be Hamiltonian.  
For example, in the case of $n=3$
we can introduce new dependent variables
$$\phi_{N+1}=v,\ \phi_{N+2}=w,\ \psi_{N+1}
=-\dfrac{1}{4}\phi_{N+2,x},\ 
\psi_{N+2}=-\dfrac{1}{4}\phi_{N+1,x}.\eqno{(2.7)}$$
The constraint (2.6) with $\beta _0=1$ becomes
$$
\left(\begin{array}{c}
c_3\\
b_3\end{array}\right)
=\dfrac{1}{4}\left(\begin{array}{c}
w_{xx}-2vw^2\\ v_{xx}-2v^2w
\end{array}\right)
=\left(\begin{array}{c}
-\phi_{N+2,x}-\dfrac{1}{2}\phi_{N+1}\phi_{N+2}^2\vspace{2mm}\\
-\phi_{N+2,x}-\dfrac{1}{2}\phi_{N+1}^2\phi_{N+2}
\end{array}\right)
=\left(\begin{array}{c}
<\psi_1,\phi_2>\\ <\psi_2,\phi_1>
\end{array}\right)
.
\eqno{(2.8)}$$
We take the Hamiltonian $H$ for the constrained system as follows
$$H=<A\psi_2,\phi_2>-<A\psi_1,\phi_1>+\phi_{N+1}<\psi_1,\phi_2>
+\phi_{N+2}<\psi_2,\phi_1>
+\dfrac{1}{4}\phi_{N+1}^2\phi_{N+2}^2-4\psi_{N+1}\psi_{N+2},$$
and thus the constrained system can be written as 
$$
\left\{\begin{array}{ll}
\phi_{1i,x}=\dfrac{\p H}
{\p \psi_{1i}}=-\lambda_i\psi_{1i}+\phi_{N+1}\phi_{2i},&
\phi_{2i,x}=\dfrac{\p H}{\p \psi_{2i}}=A\phi_{2i}+\phi_{N+2}\phi_{1i},
\vspace{2mm}\\
\phi_{N+1,x}=\dfrac{\p H}{\p \psi_{N+1}}=-4\psi_{N+2},&
\phi_{N+2,x}=\dfrac{\p H}{\p \psi_{N+2}}=-4\psi_{N+1},\vspace{2mm}\\
\psi_{1i,x}=-\dfrac{\p H}{\p \phi_{1i}}
=\lambda_i\psi_{1i}-\phi_{N+2}\psi_{2i},&
\psi_{2i,x}=-\dfrac{\p H}{\p \phi_{2i}}=
-\lambda_i\psi_{2i}-\phi_{N+1}\psi_{1i},\vspace{2mm}\\
\psi_{N+1,x}=-\dfrac{\p H}{\p \phi_{N+1}}
=-<\psi_1,\phi_2>-\dfrac{1}{2}\phi_{N+1}^2\phi_{N+2},&
\psi_{N+2,x}=-\dfrac{\p H}{\p \phi_{N+2}}
=-<\psi_2,\phi_1>-\dfrac{1}{2}\phi_{N+1}^2\phi_{N+2},
\end{array}\right.\eqno{(2.9)}$$
where $\phi_{ji}=\phi_{j}(x,\lambda_i),\ \psi_{ji}=\psi_{j}(x,\lambda_i),\
j=1,2,\,1\le i\le N$.

In general, when $n=2k+1$, the symmetry constraint (2.6) with $\beta _0=1$ is
$$\left(\begin{array}{c}
c_{2k+1}\\
b_{2k+1}\end{array}\right)=\dfrac{\delta H_{2k}}{\delta u}=
\left(\begin{array}{c}<\psi_1,\phi_2>\\
<\psi_2,\phi_1>\end{array}\right).
\eqno{(2.10)}$$
Introduce the following Jacobi-Ostrogradsky 
coordinates$^{\cite{ZengL-JMP1989}}$
$$\left.\begin{array}{lll}
\phi_{N-1+2i}
=\dfrac{d^{i-1} v}{dx^{i-1}},&
\phi_{N+2i}=\dfrac{d^{i-1} w}{dx^{i-1}},& i=1,2,...,k,\vspace{2mm}\\
\psi_{N-1+2i}=\dfrac{\delta H_{2k+1}}{\delta v^{(i)}},
&\psi_{N+2i}
=\dfrac{\delta H_{2k+1}}{\delta w^{(i)}},&
i=1,2,...,k,
\end{array}\right.$$
where $$
v^{(i)}=\dfrac{d^i v}{dx^i},\ w^{(i)}
=\dfrac{d^i w}{dx^i},\ 
\dfrac{\delta }{\delta v^{(i)}}
 =\sum_{l\geq 0}(-D)^l\dfrac{\p }{\p v^{(i+l)}} ,\ 
\dfrac{\delta }{\delta w^{(i)}}
 =\sum_{l\geq 0}(-D)^l\dfrac{\p }{\p w^{(i+l)}},\ D=\frac {d}{dx} 
,\  i\ge 0 ,$$
and denote the dependent variables by
$$\left.\begin{array}{l}
Q=(\phi_{11},...,\phi_{1N},\phi_{21},...,\phi_{2N},\phi_{N+1},...,
\phi_{N+2k})^T,\\
P=(\psi_{11},...,\psi_{1N},\psi_{21},...,\psi_{2N},\psi_{N+1},...,
\psi_{N+2k})^T. 
\end{array} \right.\eqno{(2.11)}$$
The constrained system consisting 
of (2.1) and (2.2) with all  $\lambda =\lambda _i$ and (2.10)
becomes
$$ P_x=-\dfrac{\p \~H_{2k+1}}{\p Q},\ 
 Q_x=\dfrac{\p \~H_{2k+1}}{\p P} \eqno{(2.12)}$$
with the Hamiltonian
$$ \~H_{2k+1}=
\sum^{2k}_{i=1}\phi_{N+i,x}\psi_{N+i}-H_{2k}+\hat{H}_{2k+1},$$
where the third term is given by 
$$
\hat {H}_{2k+1}=<A\psi_2,\phi_2>-<A\psi_1,\phi_1>+\phi_{N+1}<\psi_1,\phi_2>
+\phi_{N+2}<\psi_2,\phi_1>.\eqno{(2.13)}$$
When $\phi_{1i},\psi_{1i},\phi_{2i},\psi_{2i} \ ( i=1,...,N)$ 
satisfy (2.1) and (2.2) with $\lambda=\lambda_i$, we have (see (3.18) in
ref.$^{\cite{MaS-PLA1994}}$)
$$b_l=<A^{l-2k-1}\psi_2,\phi_1>,\, 
c_l=<A^{l-2k-1}\psi_1,\phi_2>,\,
a_l=\dfrac{1}{2}(<A^{l-2k-1}\psi_1,\phi_1>-<A^{l-2k-1}\psi_2,\phi_2>),
\eqno{(2.14)}$$
which are used in generating integrals of motion.
Therefore all the constrained systems possess
the Hamiltonian structures.

When $n$ is even, the regular condition for introducing the 
Jacobi-Ostrogradsky coordinates can not be satisfied . 
Nevertheless we still can introduce
new dependent variables but not the 
Jacobi-Ostrogradsky coordinates, to put the constrained system to be 
Hamiltonian. 
For example, in the case of $n=2$, since $b_2=-\dfrac{1}{2}v_x,c_2=
\dfrac{1}{2}w_x$,
the constraint (2.6) with $\beta_0=-\frac 12$ reads as
$$ v_x=<\psi_2,\phi_1>\quad w_x=-<\psi_1,\phi_2>.\eqno{(2.15)}$$
Now we do the trick by introducing new dependent variables
$$ \phi_{N+1}=v,\ \psi_{N+1}=w .$$
Two vectors of all dependent variables are given by 
$$Q=(\phi_{11},...,\phi_{1N},\phi_{21},...,\phi_{2N},
\phi_{N+1})^T,\ 
P=(\psi_{11},...,\psi_{1N},\psi_{21},...,\psi_{2N},\psi_{N+1})^T.$$
It is easy to verify that the constrained system can be transformed into
the following Hamiltonian form  
$$
\left\{\begin{array}{lll}
\phi_{1i,x}=\dfrac{\p H}{\p 
\psi_{1i}}=-\lambda_i\psi_{1i}+\phi_{N+1}\phi_{2i},&
\phi_{2i,x}=\dfrac{\p H}{\p
\psi_{2i}}=\lambda_i\phi_{2i}+\phi_{N+1}\phi_{1i},&
\phi_{N+1,x}=\dfrac{\p H}{\p \psi_{N+1}}=<\psi_2,\phi_1>,\vspace{2mm}\\
\psi_{1i,x}=-\dfrac{\p H}{\p \phi_{1i}}=\lambda_i\psi_{1i}
-\phi_{N+1}\psi_{2i},&
\psi_{2i,x}=-\dfrac{\p H}{\p \phi_{2i}}=-\lambda_i\psi_{2i}-\phi_{N+1}
\psi_{1i},&
\psi_{N+1,x}=-\dfrac{\p H}{\p \phi_{N+1}}=-<\psi_1,\phi_2>,
\end{array}\right.\eqno{(2.16)}$$
where $1\le i\le N$ and the Hamiltonian is
$$H=<A\psi_2,\phi_2>-<A\psi_1,\phi_1>+\phi_{N+1}<\psi_1,
\phi_2>+\psi_{N+1}<\psi_2,\phi_1>.$$

In the case of $n=4$,
since $b_4=\dfrac{1}{8}(-v_{xxx}+6vv_xw), c_4=\dfrac{1}{8}(w_{xxx}-6vww_x)$, 
the constraint (2.6) with $\beta_0=-\frac 12$ reads as 
$$\dfrac{1}{4}(v_{xxx}-6vwv_x)=
<\psi_2,\phi_1>,\quad\dfrac{1}{4}(w_{xxx}-6vww_x)=
-<\psi_1,\phi_2>.\eqno{(2.17)}$$
Introducing new dependent variables
$$
\left\{\begin{array}{lll}
\phi_{N+1}=v,&\phi_{N+2}=
-\dfrac{1}{2}v_x,&\phi_{N+3}=\dfrac{1}{4}v_{xx}-\dfrac{3}{8}v^2w,
\vspace{2mm}\\
\psi_{N+1}=\dfrac{1}{4}w_{xx}-\dfrac{3}{8}vw^2,&\psi_{N+2}=
\dfrac{1}{2}w_x,&\psi_{N+3}=w,
\end{array}\right.\eqno{(2.18)}$$
and two vectors of all  dependent variables
$$Q=(\phi_{11},...,\phi_{1N},\phi_{21},...,\phi_{2N},\phi_{N+1},
\phi_{N+2},\phi_{N+3})^T,\ 
P=(\psi_{11},...,\psi_{1N},\psi_{21},...,\psi_{2N},\psi_{N+1},
\psi_{N+2},\psi_{N+3})^T,$$
the corresponding constrained system becomes
$$
\left\{\begin{array}{l}
\phi_{1i,x}=-\lambda_i\psi_{1i}+\phi_{N+1}\phi_{2i},\vspace{2mm}\\
\phi_{2i,x}=\lambda_i\phi_{2i}+\phi_{N+1}\phi_{1i},\vspace{2mm}\\
\phi_{N+1,x}=-2\phi_{N+2},\vspace{2mm}\\
\phi_{N+2,x}=-2\psi_{N+3}-\dfrac{3}{4}\phi_{N+1}^2\psi_{N+3},\vspace{2mm}\\
\phi_{N+3,x}=<\psi_2,\phi_1>-\dfrac{3}{4}\phi_{N+1}^2\psi_{N+2}
-\dfrac{3}{2}\phi_{N+1}\phi_{N+2}\psi_{N+3},\vspace{2mm}\\
\psi_{1i,x}=\lambda_i\psi_{1i}-\phi_{N+3}\psi_{2i},\vspace{2mm}\\
\psi_{2i,x}=-\lambda_i\psi_{2i}-\phi_{N+1}\psi_{1i},\vspace{2mm}\\
\psi_{N+1,x}=-<\psi_1,\phi_2>+\dfrac{3}{2}\phi_{N+1}
\phi_{N+2}\phi_{N+3}+\dfrac{3}{4}\phi_{N+2}\psi_{N+3}^2,\vspace{2mm}\\
\psi_{N+2,x}=\dfrac{3}{4}\phi_{N+1}\psi_{N+3}^2+\psi_{N+1},\vspace{2mm}\\
\psi_{N+3,x}=2\psi_{N+2},
\end{array}\right.\eqno{(2.19)}$$
where $1\le i\le N$.
We can find that the Hamiltonian for this system (2.19), which
can be selected as follows
$$
\left.\begin{array}{ll}
H=&<A\psi_2,\phi_2>-<A\psi_1,\phi_1>+\phi_{N+1}<\psi_1,\phi_2>
+\psi_{N+1}<\psi_2,\phi_1>\\
&-2\phi_{N+2}\psi_{N+1}-2\phi_{N+3}\psi_{N+2}-\dfrac{3}{4}
\phi_{N+1}^2\psi_{N+2}\psi_{N+3}
-\dfrac{3}{4}\phi_{N+1}\phi_{N+2}\psi_{N+3}^2.
\end{array}\right.
\eqno{(2.20)}$$

We note that when $n=2$ or $4$, the new dependent 
variables are some kind of combinations of $b_1,
c_1$ or $b_1,b_2,b_3;c_1,c_2,c_3$. In general,
for the case of $n=2k, $ new dependent variables may similarly
be taken as combinations of
$b_1,...,b_{2k-1};c_1,...,c_{2k-1}$. However 
we do not have any concrete expression for 
such new dependent variables. How to introduce them still needs 
further investigation. 

\section{Lax representations and $r$-matrices}

For the system
$$
\left\{\begin{array}{l}
\phi_{1jx}=-\lambda_j\phi_{1j}+v\phi_{2j},\\
\phi_{2jx}=w\phi_{1j}+\lambda_j\phi_{2j},\end{array}\right.\quad
\left\{\begin{array}{l}
\psi_{1jx}=\lambda_j\psi_{1j}-w\psi_{2j},\\
\psi_{2jx}=-v\psi_{1j}-\lambda_j\psi_{2j},\end{array}\right.\quad
j=1,2,...,N,\eqno{(3.1)}
$$
we take the symmetry constraint
$$\left(\begin{array}{c}
b_n\\
c_n\end{array}\right)=\beta_0
\left(\begin{array}{c}
<\psi_2,\phi_1>\\
<\psi_1,\phi_2>
\end{array}\right)\quad (n\ge 1)\eqno{(3.2)}$$
where the $\beta_0$ is a special constant for each case of $n$.
Assume that
 $$N^{(n)}=V^{(n)}+N_0:=\left(\begin{array}{cc}
A&B\\
C&-A
\end{array}\right),\quad V^{(n)}=(\lambda^{n-1}V)_+,
\quad N_0=\beta_0\sum^N_{j=1}\left(
\begin{array}{cc}
\~a_j&\~b_j\\
\~c_j&-\~a_j\end{array}\right)\dfrac{1}
{\lambda-\lambda_j},
\eqno{(3.3)}$$
where 
$$\~a_j=\dfrac{1}{2}(\psi_{1j}\phi_{1j}
-\psi_{2j}\phi_{2j}),\quad \~b_j=\psi_{2j}\phi_{1j},\quad
\~c_j=\psi_{1j}\phi_{2j},\quad 1\le j\le N.$$
At this moment, we have
 $$N^{(n)}_x=[U,N^{(n)}],\quad U=\left(\begin{array}{cc}
 -\lambda&v\\
 w&\lambda\end{array}\right).\eqno{(3.4)}$$
The reasons are the following.
We first have 
 $$\begin{array}{l} N^{(n)}_x= V^{(n)}_x+N_{0x}=
 V^{(n)}_x+\beta_0\displaystyle{\sum^N_{j=1}\left(\begin{array}{cc}
 \~a_{jx}&\~b_{jx}\\
\~c_{jx}&-\~a_{jx}\end{array}\right)\dfrac{1}{\lambda-\lambda_j}},
\vspace{2mm}\\
\left.[U,N^{(n)}\right.]=\left.[U,V^{(n)}\right. ]
+\beta_0\displaystyle{\sum^N_{j=1}\dfrac{1}{\lambda-\lambda_j}
 \left(\begin{array}{cc}
 w\~c_j-v\~b_j&-2\~b_j\lambda-2w\~a_j\\
 2v\~a_j+2\lambda\~c_j&-w\~c_j+v\~b_j\end{array}\right)}\vspace{2mm}\\
 =[U,V^{(n)}]+\displaystyle{\beta_0\sum^N_{j=1}\dfrac{1}{\lambda-\lambda_j}}
 \left(\begin{array}{cc}
 w\~c_j-v\~b_j&-2\~b_j\lambda_j-2w\~a_j\\
 2v\~a_j+2\lambda_j\~c_j&-w\~c_j+v\~b_j\end{array}\right)
 +\beta_0\left(\begin{array}{cc}
 0&-2<\psi_2,\phi_1>\\
 2<\psi_1,\phi_2>&0\end{array}\right).\end{array} $$
Then taking the residue of (3.4) 
at $\lambda=\lambda_j$ , we get the equations
$$
\left\{\begin{array}{l}
\phi_{1j}(\psi_{2jx}+\lambda_j\psi_{2j}+v\psi_{1j})+\psi_{2j}
(\phi_{1jx}+\lambda_j\phi_{1j}-v\psi_{2j})=0,\\
\phi_{2j}(\psi_{1jx}-\lambda_j\psi_{1j}+w\psi_{2j})
+\psi_{1j}(\phi_{2jx}-\lambda_j\phi_{2j}-w\psi_{1j})=0,\\
-\phi_{2j}(\psi_{2jx}+\lambda_j\psi_{2j}+v\psi_{2j})
+\psi_{1j}(\phi_{1jx}+\lambda_j\phi_{1j}-v\psi_{2j})=0,\\
\phi_{1j}(\psi_{1jx}-\lambda_j\psi_{1j}+w\psi_{2j})
-\psi_{2j}(\phi_{2jx}-\lambda_j\phi_{2j}-w\psi_{1j})=0,
\end{array}\right.$$
which are satisfied if (3.1) holds.
The remaining relation 
$$V^{(n)}_x=[U,V^{(n)}]+\beta_0\left(\begin{array}{cc}
0&-2<\psi_2,\phi_1>\\
2<\psi_1,\phi_2>&0\end{array}\right)$$
is nothing but the constraint (3.2).
Therefore the equation (3.4) holds if the system (3.1) holds together with
the symmetry constraint (3.2). 
This means that $N^{(n)}$ and $U$
constitutes a Lax pair 
for the constrained system consisting of (3.1) and (3.2).

Lax operators $N^{(n)}$ can be represented explicitly. For example,
in the case of $n=1$, we have
$$\left\{\begin{array}{l}A(\lambda)=-1+\sum^N_{j=1}
\dfrac{\beta_0}{\lambda-\lambda_j}\dfrac{1}{2}
(\psi_{1j}\phi_{1j}-\psi_{2j}\phi_{2j}):= -1+A_0,\vspace{2mm}\\
B(\lambda)=\sum^N_{j=1}\dfrac{\beta_0}{\lambda-
\lambda_j}\psi_{2j}\phi_{1j}:= B_0,\vspace{2mm}\\
C(\lambda)=\sum^N_{j=1}\dfrac{\beta_0}{\lambda-
\lambda_j}\psi_{1j}\phi_{2j}:= C_0.
\end{array}\right.
\eqno{(3.5)}$$
The Poisson bracket in this case is given by 
$$
\{f,g\}=\sum^2_{i=1}\sum^N_{j=1}(\dfrac{\p f}
{\p \psi_{ij}}\dfrac{\p g}{\p \phi_{ij}}-\dfrac{\p f}{\p \phi_{ij}}
\dfrac{\p g}{\p \psi_{ij}}),\eqno{(3.6)}$$
as defined in ref.$^{\cite{MaS-PLA1994}}$. 
For each case of $n$, we use the standard 
Poisson bracket like the above.

Now turn to constructing $r$-matrices$^{\cite{FaddeevT-BookHMST1987}}$. 
A direct calculation, similar to 
refs.$^{\cite{ZengH-JPA1996},\cite{Zeng-IP1997}}$,
gives rise to
$$\left\{\begin{array}{l}
\{A(\lambda),A(\mu)\}=\{B(\lambda),B(\mu)\}=\{C(\lambda),C(\mu)\}=0,\\
\{A(\lambda),B(\mu)\}=\dfrac{\beta_0}{\lambda-\mu}(B(\mu)-B(\lambda)),\\
\{A(\lambda),C(\mu)\}=\dfrac{\beta_0}{\lambda-\mu}(C(\lambda)-C(\mu)),\\
\{B(\lambda),C(\mu)\}=\dfrac{2\beta_0}{\lambda-\mu}(A(\mu)-A(\lambda)),
\end{array}\right.\eqno{(3.7)}$$
for $A(\lambda ),B(\lambda ),C(\lambda )$ defined by (3.5).
Setting $$M(\lambda)=\left(\begin{array}{cc}
A(\lambda)&B(\lambda)\\
C(\lambda)&-A(\lambda)\end{array}\right),\quad M_1=M\otimes I,
\  M_2=I\otimes M ,\ 
P=\left(\begin{array}{cccc}
1&0&0&0\\
0&0&1&0\\
0&1&0&0\\
0&0&0&1\end{array}\right),
\eqno{(3.8)}$$
we have
$$
\{M_1(\lambda), M_2(\mu)\}=\left(\begin{array}{cccc}
\{A(\lambda),A(\mu)\}&\{A(\lambda),B(\mu)\}&\{B(\lambda),A(\mu)\}&
\{B(\lambda),B(\mu)\}\\
\{A(\lambda),C(\mu)\}&\{A(\lambda),-A(\mu)\}&\{B(\lambda),
C(\mu)\}&\{B(\lambda),-A(\mu)\}\\
\{C(\lambda),A(\mu)\}&\{C(\lambda),B(\mu)\}&\{-A(\lambda),
A(\mu)\}&\{-A(\lambda),B(\mu)\}\\
\{C(\lambda),C(\mu)\}&\{C(\lambda),-A(\mu)\}&\{-A(\lambda),
-A(\mu)\}&\{-A(\lambda),-A(\mu)\}
\end{array}\right),\eqno{(3.9)}$$
$$M_1(\lambda)+M_2(\mu)=\left(\begin{array}{cccc}
A(\lambda)+A(\mu)&B(\mu)&B(\lambda)&0\\
C(\mu)&A(\lambda)-A(\mu)&0&B(\lambda)\\
C(\lambda)&0&-A(\lambda)+A(\mu)&B(\mu)\\
0&C(\lambda)&C(\mu)&-A(\lambda)-A(\mu)\end{array}\right),
\eqno{(3.10)}$$
$$[P,M_1(\lambda)+M_2(\mu)]=\left(\begin{array}{cccc}
0&B(\mu)-B(\lambda)&B(\lambda)-B(\mu)&0\\
C(\lambda)-C(\mu)&0&2(-A(\lambda)+A(\mu))&B(\mu)-B(\lambda)\\
C(\mu)-C(\lambda)&2(A(\lambda)-A(\mu))&0&B(\lambda)-B(\mu)\\
0&C(\lambda)-C(\mu)&C(\mu)-C(\lambda)&0\end{array}\right).\eqno{(3.11)}$$
Comparing (3.9) with (3.11) and using (3.7), we obtain the classical
 Poisson structure$^{\cite{FaddeevT-BookHMST1987}}$
$$\{M_1(\lambda), M_2(\mu)\}=[R(\lambda ,\mu),
M_1(\lambda)+M_2(\mu)]\eqno{(3.12)}$$
with the following $r$-matrix
$$R(\lambda,\mu)=\dfrac{\beta_0}{\lambda-\mu}P.\eqno{(3.13)}$$
It follows from (3.12) that$^{\cite{ZengH-JPA1996}-\cite{BabelonV-PLB1990}}$
$$\left.\begin{array}{l}
\{M_1^2(\lambda), M_2^2(\mu)\}=[R_1(\lambda,\mu), M_1(\lambda)+M_2(\mu)],
\vspace{2mm}\\
R_1(\lambda,\mu)=\sum^1_{i=0}
\sum^1_{j=0}M_1(\lambda)^{1-i}M_2(\mu)^{1-j}
R(\lambda,\mu)M_1^i(\lambda)M_2^j(\mu).
\end{array}\right.\eqno{(3.14)}$$
This equality implies the relation
$$\{{\rm Tr}M^2(\lambda),{\rm Tr}M^2(\mu)\}
=0.$$
Therefore 
the integrals of motion generated from ${\rm Tr}M^2$ are in involution in 
pairs.

Let us now show more examples.
For $n=2$, we have
$$A(\lambda)=-\lambda+A_0,\quad 
B(\lambda)=\phi_{N+1}+B_0,\quad C(\lambda)=\psi_{N+1}+C_0.\eqno{(3.15)}$$
For $n=3$, we have
$$\left\{\begin{array}{l}
A(\lambda)=-\lambda^2+\dfrac{1}{2}\phi_{N+1}\phi_{N+2}+A_0,
\vspace{2mm}\\
 B(\lambda)=\phi_{N+1}\lambda+2\psi_{N+2}+B_0,
\vspace{2mm}\\ C(\lambda)=
\phi_{N+2}\lambda-2\psi_{N+1}+C_0.
\end{array}\right.
\eqno{(3.16)}$$
For $n=4$, we have
$$\left\{\begin{array}{l}
A(\lambda)=-\lambda^3+\dfrac{1}{2}\phi_{N+1}\psi_{N+3}
\lambda+\dfrac{1}{2}\phi_{N+1}\psi_{N+2}+\dfrac{1}{2}\psi_{N+3}\phi_{N+2}
+A_0,\vspace{2mm}\\
B(\lambda)=\phi_{N+1}\lambda^2+\phi_{N+2}\lambda+\phi_{N+3}
-\dfrac{1}{8}\phi_{N+1}^2\psi_{N+3}+B_0,\vspace{2mm}\\
C(\lambda)=\phi_{N+3}\lambda^2+\psi_{N+2}\lambda+\psi_{N+1}
-\dfrac{1}{8}\phi_{N+1}\psi_{N+3}^2+C_0.
\end{array}\right.
\eqno{(3.17)}$$

In the case of $n=3$, the commutator relation (3.7) 
is established only for $\beta_0=1$.
In the cases of $n=2$ and $n=4$, 
the commutator relation (3.7) is established only for $\beta_0=-\dfrac{1}{2}$.
All above result for $n=1$ can be extended to the cases of $n=2,3,4$.
Therefore all corresponding constrained systems have 
the $r$-matrices determined by (3.13).

\section{Integrals of motion and Liouville integrability}

First of all, we point out that if $\phi_{ij},\psi_{ij},
\,i=1,2\,(1\le j\le N),$ are solutions
of (3.1) with $\lambda=\lambda_j$, then 
a direct calculation yields$^{\cite{MaS-PLA1994},\cite{MaFO-PA1996}}$ that
$$
\dfrac{d}{dx}\~F_i=0,\quad \~F_j=\psi_{1j}\phi_{1j}+\psi_{2j}\phi_{2j}.
\eqno{(4.1)}$$

From the relation $$\dfrac{d}{dx}\sum_{m=0}^{\infty} F_{m+(n-1)}
\lambda ^{-m+2(n-1)}:=\frac12 \dfrac{d}{dx}{\rm Tr}(N^{(n)})^2=
\frac12 \dfrac{d}{dx}{\rm Tr}
\left(\begin{array}{cc}
A(\lambda ) &B(\lambda ) \\C(\lambda ) &-A(\lambda ) 
\end{array}\right)^2=0,$$ 
which may be resulted 
directly from the Lax representation $N^{(n)}_x=[U,N^{(n)}]$,
we can get another set of integrals of motion $\{F_m\}_n^\infty$ for 
each case of $n$. Now we
list them as follows.

i) $n=1$. The corresponding constraint is 
$v=<\psi_2,\phi_1>,\ w=<\psi_1,\phi_2>$, and the integrals of motion
$F_n$ has been found in ref.$^{[10]}$
$$\left.\begin{array}{l}
F_1=<\psi_2,\phi_2>-<\psi_1,\phi_1>,\\
F_n=\sum^n_{i=1}(a_ia_{n-i}+b_ic_{n-i})-2a_n,\quad n\geq 2,
\end{array}\right.$$
where $$
a_l=\dfrac{1}{2}(<A^{l-1}\psi_1,\phi_1>-<A^{l-1}\psi_2,\phi_2>), 
\ b_l=<A^{l-1}\psi_2,\phi_1>,\ c_l=<A^{l-1}\psi_1,\phi_2>.$$
By using the idea in Refs.\cite{MaDZL-NCB1996} \cite{MaFO-PA1996},
the required $2N$ integrals of motion for
Liouville integrability of
the constrained system in the case of
$n=1$ can be chosen as  
$$\~F_1,\~F_2,...,\~F_N,F_1,...,F_N.\eqno{(4.2)}$$

ii) $n=2$. The corresponding
constraint is $v_x=<\psi_2,\phi_1>,\ w_x=-<\psi_1,\phi_2>$, and
similarly the required $2N+1$ integrals of motion can be chosen as
$$\~F_1,\~F_2,...,\~F_N,F_2,F_3,...,F_{N+2},\eqno{(4.3)}$$
where 
$$\left.\begin{array}{ll}
F_2=&\dfrac{1}{2}(<\psi_1,\phi_1>-<\psi_2,\phi_2>)+\phi_{N+1}\psi_{N+1},
\vspace{2mm}\\
F_3=&\dfrac{1}{2}(<A\psi_1,\phi_1>-<A\psi_2,\phi_2>)-\dfrac{\phi_{N+1}}{2}
<\psi_1,\phi_2>
-\dfrac{\psi_{N+1}}{2}<\psi_2,\phi_1>,\quad H=-\dfrac{1}{2}F_3,\vspace{2mm}\\
F_n=&\dfrac{1}{2}(<A^{n-2}\psi_1,\phi_1>-<A^{n-2}\psi_2,\phi_2>)
-\dfrac{\phi_{N+1}}{2}<A^{n-3}\psi_1,\phi_2>
-\dfrac{\psi_{N+1}}{2}<A^{n-3}\psi_2,\phi_1>\\
&+\displaystyle{\sum_{l+k=n,k,l\geq 2}(a_la_k+b_lc_k) },\quad
 n\geq 4,\vspace{2mm}\\
a_l=&-\dfrac{1}{4}(<A^{l-2}\psi_1,\phi_1>-<A^{l-2}\psi_2,\phi_2>),\ 
b_l=-\dfrac{1}{2}<A^{l-2}\psi_2,\phi_1>,\  c_l=-\dfrac{1}{2}<A^{l-2}
\psi_1,\phi_2>.\end{array}\right.$$ 

iii) $n=3$. The corresponding
constraint is $\dfrac{1}{4}(v_{xx}-2v^2w)=<\psi_2,\phi_1>,\ 
\dfrac{1}{4}(w_{xx}-2vw^2)=<\psi_1,\phi_2>$.
The required $2N+2$ integrals of motion can be chosen as
$$\~F_1,\~F_2,...,\~F_N,F_3,F_4,...,F_{N+4},\eqno{(4.4)}$$
where
$$\left.\begin{array}{ll}
F_3=&-<\psi_1,\phi_1>+<\psi_2,\phi_2>+2\psi_{N+2}
\phi_{N+2}-2\psi_{N+1}\phi_{N+1},\vspace{2mm}\\
F_4=&-<A\psi_1,\phi_1>+<A\psi_2,\phi_2>+\phi_{N+1}<\psi_1,\phi_2>+\phi_{N+2}
<\psi_2,\phi_1>+
\dfrac{1}{4}\phi_{N+1}^2\phi_{N+2}^2-4\psi_{N+1}\psi_{N+2}=H,\vspace{2mm}\\
F_5=&-<A^{2}\psi_1,\phi_1>+<A^{2}\psi_2,\phi_2>
+\phi_{N+1}<A\psi_1,\phi_2>+\phi_{N+2}<A\psi_2,\phi_1>\vspace{2mm}
\\
&+2\psi_{N+2}<\psi_1,\phi_2>
-2\psi_{N+1}<\psi_2,\phi_1>
+\dfrac{1}{2}\phi_{N+1}\phi_{N+2}(<\psi_1,\phi_1>
 -<\psi_2,\phi_2>),
\vspace{2mm}\\
F_n=&-<A^{n-3}\psi_1,\phi_1>+<A^{n-3}\psi_2,\phi_2>
+\phi_{N+1}<A^{n-4}\psi_1,\phi_2>+\phi_{N+2}<A^{n-4}\psi_2,\phi_1>\vspace{2mm}
\\
&+2\psi_{N+2}<A^{n-5}\psi_1,\phi_2>-2\psi_{N+1}<A^{n-5}\psi_2,\phi_1>
+\dfrac{1}{2}\phi_{N+1}\phi_{N+2}(<A^{n-5}\psi_1,\phi_1>
\vspace{2mm}\\& -<A^{n-5}\psi_2,\phi_2>)
+\displaystyle{\sum_{l+k=n,k,l\geq 3}(a_la_k+b_lc_k)},\quad  n\geq 6,
\vspace{2mm}\\
a_l=&\dfrac{1}{2}(<A^{l-3}\psi_1,\phi_1>-<A^{l-3}\psi_2,\phi_2>),\ 
b_l=<A^{l-3}\psi_2,\phi_1>,\  c_l=<A^{l-3}\psi_1,\phi_2>.
\end{array}\right.$$

iii) $n=4$. The corresponding 
constraint is $\dfrac{1}{4}(v_{xxx}-6vwv_x)=<\psi_2,\phi_1>,\ 
\dfrac{1}{4}(w_{xxx}-6vww_x)=-<\psi_1,\phi_2>$.
The required $2N+3$ integrals of motion may be chosen as
$$\~F_1,\~F_2,...,\~F_N,F_4,F_5,F_6,...,F_{N+6},\eqno{(4.5)}$$
where 
$$\left.\begin{array}{ll}
F_4=&\dfrac{1}{2}(<\psi_1,\phi_1>-<\psi_2,\phi_2>)
+\dfrac{1}{4}\phi_{N+1}^2\psi_{N+3}^2
+\phi_{N+1}\psi_{N+1}+\phi_{N+3}\psi_{N+3}+\phi_{N+2}\psi_{N+2},\vspace{2mm}\\
F_5=&\dfrac{1}{2}(<A\psi_1,\phi_1>-<A\psi_2,\phi_2>)
-\dfrac{\phi_{N+1}}{2}<\psi_1,\phi_2>-\dfrac{\phi_{N+3}}{2}<\psi_2,\phi_1>
+\phi_{N+2}\psi_{N+1}\vspace{2mm}\\ &+\phi_{N+3}\psi_{N+2}
+\dfrac{3}{8}\phi_{N+1}\phi_{N+2}\psi_{N+3}^2+\dfrac{3}{8}
\phi_{N+1}^2\psi_{N+2}\psi_{N+3},\quad H=-\dfrac{1}{2}F_5,\vspace{2mm}\\
F_6=&\dfrac{1}{2}(<A^2\psi_1,\phi_1>-<A^2\psi_2,\phi_2>)
-\dfrac{\phi_{N+1}\psi_{N+3}}{4}(<\psi_1,\phi_1>-<\psi_2,\phi_2>)
-\dfrac{\phi_{N+1}}{2}<A\psi_1,\phi_2>\vspace{2mm}\\
&-\dfrac{\psi_{N+3}}{2}<A\psi_2,\phi_1>-\dfrac{\phi_{N+2}}{2}
<\psi_1,\phi_2>-\dfrac{\psi_{N+2}}{2}<\psi_2,\phi_1>+\dfrac{1}{4}
(\phi_{N+1}^2\psi_{N+2}^2+\phi_{N+2}^2\psi_{N+3}^2\vspace{2mm}\\
&+2\phi_{N+1}\psi_{N+2}\psi_{N+3}\phi_{N+2})+\phi_{N+3}\psi_{N+1}
-\dfrac{1}{8}\phi_{N+3}\phi_{N+1}\psi_{N+3}^2-\dfrac{1}{8}
\phi_{N+1}^2\psi_{N+1}\psi_{N+3}+\dfrac{1}{64}\phi_{N+1}^3\psi_{N+3}^3,
\vspace{2mm}\\
F_7=&\dfrac{1}{2}(<A^{3}\psi_1,\phi_1>-<A^{3}\psi_2,\phi_2>)
-\dfrac{1}{4}\phi_{N+1}\psi_{N+3}(<A\psi_1,\phi_1>-<A
\psi_2,\phi_2>)\vspace{2mm}\\
&-\dfrac{1}{4}(\phi_{N+1}\psi_{N+1}+\psi_{N+3}\psi_{N+2})
(<\psi_1,\phi_1>-
<\psi_2,\phi_2>)-\dfrac{\phi_{N+1}}{2}<A^2\psi_1,\phi_2>
\vspace{2mm}\\
&-\dfrac{\psi_{N+3}}{2}<A^{2}\psi_2,\phi_1>-\dfrac{\phi_{N+2}}{2}
<A\psi_1,\phi_2>-\dfrac{\psi_{N+2}}{2}<A\psi_2,\phi_1>\vspace{2mm}\\&
-\dfrac{1}{2}(\phi_{N+3}
-\dfrac{1}{8}\phi_{N+1}^2\psi_{N+3})<\psi_1,\phi_2>
-\dfrac{1}{2}(\psi_{N+1}-\dfrac{1}{8}\phi_{N+1}\psi_{N+3}^2)
 <\psi_2,\phi_1>,
\vspace{2mm}\\
F_n=&\dfrac{1}{2}(<A^{n-4}\psi_1,\phi_1>-<A^{n-4}\psi_2,\phi_2>)
-\dfrac{1}{4}\phi_{N+1}\psi_{N+3}(<A^{n-6}\psi_1,\phi_1>-<A^{n-6}
\psi_2,\phi_2>)\vspace{2mm}\\
&-\dfrac{1}{4}(\phi_{N+1}\psi_{N+1}+\psi_{N+3}\psi_{N+2})
(<A^{n-7}\psi_1,\phi_1>-
<A^{n-7}\psi_2,\phi_2>)-\dfrac{\phi_{N+1}}{2}<A^{n-5}\psi_1,\phi_2>
\vspace{2mm}\\
&-\dfrac{\psi_{N+3}}{2}<A^{n-5}\psi_2,\phi_1>-\dfrac{\phi_{N+2}}{2}
<A^{n-6}\psi_1,\phi_2>-\dfrac{\psi_{N+2}}{2}<A^{n-6}\psi_2,\phi_1>
\vspace{2mm}\\
&-\dfrac{1}{2}(\phi_{N+3}
-\dfrac{1}{8}\phi_{N+1}^2\psi_{N+3})<A^{n-7}\psi_1,\phi_2>
-\dfrac{1}{2}(\psi_{N+1}-\dfrac{1}{8}\phi_{N+1}\psi_{N+3}^2)
 <A^{n-7}\psi_2,\phi_1>\vspace{2mm}\\
&+\displaystyle{\sum_{l+k=n,k,l\geq 4}(a_la_k+b_lc_k)},\quad  n\geq 8,
\vspace{2mm}\\
a_l=&-\dfrac{1}{4}(<A^{l-4}\psi_1,\phi_1>-<A^{l-4}\psi_2,\phi_2>),
\ b_l=-\dfrac{1}{2}<A^{l-4}\psi_2,\phi_1>, \ 
c_l=-\dfrac{1}{2}<A^{l-4}\psi_1,\phi_2>.
\end{array}\right.$$

The involution property 
$\{\~F_i, \~F_j\}=\{\~F_i, F_j\}=0$ can be easily checked
and the involution property $\{F_i,F_j\}=0$ is a
direct conclusion from the $r$-matrices in $\S3$. 
Now the remaining problem for Liouville integrability is 
to prove that all the required integrals of motion are functionally 
independent. This can be achieved, indeed. The following analysis  
aims to providing a mathematical proof for it. 

i) The case of $n=1$. Assume that the result
on the functional independence is not true.
Then there exist $2N$ integrals
$\alpha_1,...\alpha_N,\beta_1,...\beta_N,$ satisfying
$\sum^N_{i=1}\alpha_i^2+\sum^N_{j=1}\beta_j^2\not=0$, such that
$$\sum^N_{i=1}\alpha_i\~F_i+\sum^N_{j=1}\beta_jF_j=0.$$
Since $\dfrac{\p \~F_j}{\p \phi_{il}}=\psi_{ij}\delta_{jl}, 
 \dfrac{\p F_n}{\p \phi_{1l}}|_{\phi_{1}=\phi_{2}=0}=
-\lambda_l^{n-1}\psi_{1l},
\dfrac{\p F_n}{\p \phi_{2l}}=\lambda_l^{n-1}\psi_{2l},\ n\ge 1,$ we have
$$\left.\begin{array}{l}
\Delta :=\textrm{det}\left(\begin{array}{cccccc}
\dfrac{\p \~F_1}{\p \phi_1}&\cdots &\dfrac{\p \~F_N}{\p \phi_1}
&\dfrac{\p F_1}{\p \phi_1}&\cdots &\dfrac{\p F_N}{\p \phi_1}\\
\dfrac{\p \~F_1}{\p \phi_2}&\cdots &  \dfrac{\p \~F_N}{\p \phi_2}
&\dfrac{\p F_1}{\p \phi_2} &\cdots &  \dfrac{\p F_N}{\p \phi_2}
\end{array}\right)_{\phi_{1}=\phi_{2}=0}
\vspace{2mm}\\
=\textrm{det}
\left(\begin{array}{cccccc}
\psi_{11}& &0&-\psi_{11}&\cdots&-\lambda_1^{N-1}\psi_{11}\\
 &\ddots&&\vdots&&\vdots\\
0&&\psi_{1N}&-\psi_{1N}&\cdots&-\lambda_N^{N-1}\psi_{1N}\\
\psi_{21}& &0&\psi_{21}&\cdots&\lambda_1^{N-1}\psi_{21}\\
 &\ddots&&\vdots&&\vdots\\
0&&\psi_{2N}&\psi_{2N}&\cdots&\lambda_N^{N-1}\psi_{2N}
\end{array}\right)
\vspace{2mm}\\
=2^N\prod ^2_{i=1}\prod ^N_{j=1}\psi_{ij}
\left|\begin{array}{cccc}
1&\lambda_1&\cdots&\lambda_1^{N-1}\\
\vdots&\vdots&&\vdots\\
1&\lambda_N&\cdots&\lambda_N^{N-1}
\end{array}\right|
\not\equiv 0.
\end{array}\right.$$
This means that all $\alpha_i,\beta_j$ must be zero. Therefore the functions
$\~F_k,F_k(1\leq k\leq N)$ can be functionally independent at least on certain
region of $R^{2N}$.

ii) The case of $n=2$. We need to show that
the required $2N+1$ integrals of motion are functionally independent.
Since we have 
$$\begin{array}{l}
\dfrac{\p F_2}{\p \phi_{1l}} =\dfrac12 \psi_{1l},\
\dfrac{\p F_2}{\p \phi_{2l}} =-\dfrac12 \psi_{2l},
\vspace{2mm}\\
\dfrac{\p F_n}{\p \phi_{1l}}|_{\phi_{1}=\phi_{2}=\phi_{N+1}=0}=
\dfrac{1}{2}(\lambda_l^{n-2}\psi_{1l}-\lambda_l^{n-3}\psi_{N+1}\psi_{2l})
=\dfrac{1}{2}\lambda_l^{n-3}\psi_{1l,x},\ n\ge 3,
\vspace{2mm}\\
\dfrac{\p F_n}{\p \phi_{2l}}|_{\phi_{1}=\phi_{2}=\phi_{N+1}=0}=
-\dfrac{1}{2}\lambda_l^{n-2}\psi_{2l},\ n\ge 3, \vspace{2mm}\\
\dfrac{\p F_2}{\p \phi_{N+1}}|_{\phi_{1}=\phi_{2}=\phi_{N+1}=0}=
\psi_{N+1},\ 
\dfrac{\p F_n}{\p \phi_{N+1}}|_{\phi_{1}=\phi_{2}=\phi_{N+1}=0}=
0,\ n\ge 3,\end{array}$$
we can make the following calculation
 $$\left.
 \begin{array}{l}
 \Delta:=\textrm{det}\left(\begin{array}{cccccc}
\dfrac{\p \~F_1}{\p \phi_1} &\cdots&   \dfrac{\p \~F_N}{\p \phi_1}
&\dfrac{\p F_2}{\p \phi_1}&\cdots&   \dfrac{\p F_{N+2}}{\p \phi_1}\\
\dfrac{\p \~F_1}{\p \phi_2} &\cdots&   \dfrac{\p \~F_N}{\p \phi_2}
&\dfrac{\p F_2}{\p \phi_2}&\cdots&   \dfrac{\p F_{N+2}}{\p \phi_2}\\
\dfrac{\p \~F_1}{\p \phi_{N+1}}&\cdots& \dfrac{\p \~F_N}{\p \phi_{N+1}} 
&\dfrac{\p F_2}{\p \phi_{N+1}} &\cdots& \dfrac{\p F_{N+2}}{\p \phi_{N+1}}
\end{array}\right)_{\phi_{1}=\phi_{2}=\phi_{N+1}=0}
\vspace{2mm}\\
=\textrm{det}\left(\begin{array}{ccccccc}
\psi_{11}& &0&\dfrac{1}{2}\psi_{11}&\dfrac{1}{2}\psi_{11,x}
&\cdots&\dfrac{1}{2}\lambda_1^{N-1}\psi_{11,x}\\
 &\ddots&&\vdots&\vdots&&\vdots\\
0&&\psi_{1N}&\dfrac{1}{2}\psi_{1N}&\dfrac{1}{2}\psi_{1N,x}
&\cdots&\dfrac{1}{2}\lambda_N^{N-1}\psi_{1N,x}\\
\psi_{21}& &0&-\dfrac{1}{2}\psi_{21}&-\dfrac{1}{2}\lambda_1\psi_{21}
&\cdots&-\dfrac{1}{2}\lambda_1^N\psi_{21}\\
 &\ddots&&\vdots&\vdots&&\vdots\\
0&&\psi_{2N}&-\dfrac{1}{2}\psi_{2N}&-\dfrac{1}{2}\lambda_{N}\psi_{2N}
&\cdots&-\dfrac{1}{2}\lambda_N^N\psi_{2N}\\
0&\cdots&0&\psi_{N+1}&0&\cdots&0
\end{array}\right)
\vspace{2mm}\\
=(-1)^{N}\psi_{N+1}
\left|\begin{array}{cccccc}
\psi_{11}&&0
&\dfrac{1}{2}\psi_{11,x}&\cdots&\dfrac{1}{2}\lambda_1^{N-1}\psi_{11,x}\\
&\ddots&&\vdots&&\vdots\\
0&&\psi_{1N}&
\dfrac{1}{2}\psi_{1N,x}&\cdots&\dfrac{1}{2}\lambda_1^{N-1}\psi_{1N,x}\\
\psi_{21}&&0&
-\dfrac{1}{2}\lambda_1\psi_{21}&\cdots&-\dfrac{1}{2}\lambda_1^N\psi_{21}
\\ &\ddots&&\vdots&&\vdots\\
0&&\psi_{2N}&
-\dfrac{1}{2}\lambda_N\psi_{2N}&\cdots&-\dfrac{1}{2}\lambda_N^N\psi_{2N}
\end{array}\right|
\vspace{2mm}\\
=\psi_{N+1}\prod _{j=1}^N\psi _{2j} 
\left|\begin{array}{ccc}
\frac 12 \lambda _1\psi_{11}+\frac1 2 \psi_{11,x}&\cdots &
\frac 12 \lambda _1^N\psi_{11}+\frac1 2  \lambda _1^{N-1}\psi_{11,x}\\
\vdots &\ddots &\vdots\\
\frac 12 \lambda _N\psi_{1N}+\frac1 2 \psi_{1N,x}&\cdots &
\frac 12 \lambda _N^N\psi_{1N}+\frac1 2  \lambda _N^{N-1}\psi_{1N,x}
\end{array}\right|
\vspace{2mm}\\
=(\dfrac 12 )^N\psi_{N+1}
\prod ^N_{j=1}\psi_{2j}(\lambda_j\psi_{1j}+\psi_{1j,x})
\left|\begin{array}{cccc}
1&\lambda_1&\cdots&\lambda_1^{N-1}\\
\vdots&\vdots&&\vdots\\
1&\lambda_N&\cdots&\lambda_N^{N-1}
\end{array}\right|\not\equiv 0.
\end{array}\right.$$
Therefore the required integrals of motion are functionally independent
at least on one region.

iii) The case of $n=3$. The required  $2N+2$ integrals of motion
are functionally independent, because we have
$$\left.\begin{array}{l}
\Delta:=\left(\begin{array}{cccccc}
\dfrac{\p \~F_1}{\p \phi_1} &\cdots & \dfrac{\p \~F_N}{\p \phi_1}
&\dfrac{\p F_3}{\p \phi_1} &\cdots & \dfrac{\p F_{N+4}}{\p \phi_1}\\
\dfrac{\p \~F_1}{\p \phi_2} &\cdots & \dfrac{\p \~F_N}{\p \phi_2}
&\dfrac{\p F_3}{\p \phi_2}&\cdots & \dfrac{\p F_{N+4}}{\p \phi_2}\\
\dfrac{\p \~F_1}{\p \psi_{N+1}}&\cdots & \dfrac{\p \~F_N}{\p \psi_{N+1}}
&\dfrac{\p F_3}{\p \psi_{N+1}}&\cdots & \dfrac{\p F_{N+4}}{\p \psi_{N+1}}\\
\dfrac{\p \~F_1}{\p \phi_{N+2}}&\cdots & \dfrac{\p \~F_N}{\p \phi_{N+2}}
&\dfrac{\p F_3}{\p \phi_{N+2}}&\cdots & \dfrac{\p F_{N+4}}{\p \phi_{N+2}}
\end{array}\right)_{\scriptsize{\begin{array}{l}
\phi_{1}=\phi_{2}=0\\
\phi_{N+1}=\phi_{N+2}=\psi_{N+1}=0\end{array}}}
\vspace{2mm}\\
=\textrm{det}
\left(\begin{array}{cccccccc}
\psi_{11}& &0& -\psi_{11}&
-\lambda_1\psi_{11}&-\lambda_1^2\psi_{11}&\cdots&-\lambda_1^{N+1}\psi_{11}\\
 &\ddots&& \vdots&\vdots&\vdots&&\vdots\\
0&&\psi_{1N}& -\psi_{1N}&
-\lambda_N\psi_{1N}&-\lambda_N^2\psi_{1N}&\cdots&-\lambda_N^{N+1}\psi_{1N}\\
\psi_{21}& &0& \psi_{21}&\lambda_1\psi_{21}
 &\lambda_1^2\psi_{21} +2\psi_{N+2}\psi_{11}&\cdots&
 \lambda_1^{N+1}\psi_{21}+2\lambda_1^{N-1}\psi_{N+2}\psi_{11}\\
 &\ddots&& \vdots&\vdots&\vdots&&\vdots\\
0& &\psi_{2N}& \psi_{2N}&
\lambda_N\psi_{2N}&\lambda_N^2\psi_{2N}+2\psi_{N+2}\psi_{1N}&\cdots&
\lambda_N^{N+1}\psi_{2N}+2\lambda_N^{N-1}\psi_{N+2}\psi_{1N}
\\
 0&\cdots&0&0&-4\psi_{N+2}&0&\cdots&0\\
0&\cdots&0& 2\psi_{N+2}&0&0&\cdots&0
\end{array}\right)
\vspace{2mm}\\
=8\psi_{N+2}^2\left|\begin{array}{cccccc}
\psi_{11}&\cdots &0& -\lambda _1^2\psi_{11}&\cdots &
-\lambda _{1}^{N+1}\psi_{11}\\
&\ddots & &
\vdots & & \cdots \\
0& &\psi_{1N}&
-\lambda _N^2\psi_{1N}&\cdots &
-\lambda _{N}^{N+1}\psi_{1N}\\
\psi_{21}&\cdots &0& \lambda _1^2\psi_{21}+2\psi_{N+2}\psi_{11}&\cdots &
\lambda _{1}^{N+1}\psi_{21}+2\lambda _1^{N-1}\psi_{N+2}\psi_{1N}\\
&\ddots & &
\vdots & & \cdots \\
0& &\psi_{2N}&
 \lambda _N^2\psi_{2N}+2\psi_{N+2}\psi_{1N}&\cdots &
\lambda _{N}^{N+1}\psi_{2N}+2\lambda _N^{N-1}\psi_{N+2}\psi_{1N}
\end{array}\right|
\vspace{2mm}\\
=2^{N+3}\psi_{N+2}^2\prod ^N_{j=1}\psi_{1j}(\lambda_j^2\psi_{2j}+\psi_{N+2}
\psi_{1j})
\left|\begin{array}{cccc}
1&\lambda_1&\cdots&\lambda_1^{N-1}\\
\vdots&\vdots&&\vdots\\
1&\lambda_N&\cdots&\lambda_N^{N-1}
\end{array}\right|\not\equiv 0.
\end{array}\right.$$

iv) The case of $n=4$.
The required $2N+3$ integrals of motion are functionally independent, because
we have
 $$\left.\begin{array}{l}
 \Delta:=\textrm{det}\left(\begin{array}{cccccc}
\dfrac{\p \~F_1}{\p \phi_1}  &\cdots & \dfrac{\p \~F_N}{\p \phi_1}
&\dfrac{\p F_4}{\p \phi_1} &\cdots & \dfrac{\p F_{N+6}}{\p \phi_1}\\
\dfrac{\p \~F_1}{\p \phi_2} &\cdots & \dfrac{\p \~F_N}{\p \phi_2}
&\dfrac{\p F_4}{\p \phi_2} &\cdots & \dfrac{\p F_{N+6}}{\p \phi_2}\\
\dfrac{\p \~F_1}{\p \psi_{N+1}} &\cdots & \dfrac{\p \~F_N}{\p \psi_{N+1}}
&\dfrac{\p F_4}{\p \psi_{N+1}}&\cdots & \dfrac{\p F_{N+6}}{\p \psi_{N+1}}\\
\dfrac{\p \~F_1}{\p \psi_{N+2}}&\cdots &\dfrac{\p \~F_{N}}{\p \psi_{N+2}}
&\dfrac{\p F_4}{\p \psi_{N+2}}&\cdots &\dfrac{\p F_{N+6}}{\p \psi_{N+2}}\\
\dfrac{\p \~F_1}{\p \psi_{N+3}} &\cdots &\dfrac{\p \~F_N}{\p \psi_{N+3}} 
&\dfrac{\p F_4}{\p \psi_{N+3}}&\cdots &\dfrac{\p F_{N+6}}{\p \psi_{N+3}} 
\end{array}\right)_{\scriptsize{\begin{array}{l}
\phi_{1}=\phi_{2}=0\\
\phi_{N+1}=\phi_{N+2}=0\\
\psi_{N+1}=\psi_{N+2}=\psi_{N+3}=0\end{array}}}\vspace{2mm}\\
=\textrm{det}\left(\begin{array}{cccccc}
\psi_{11}& &0
&\dfrac{1}{2}\psi_{11}&\dfrac{1}{2}\lambda_1\psi_{11}
&\dfrac{1}{2}\lambda_1^2\psi_{11}
\\
 &\ddots& &\vdots &\vdots &\vdots
 \\
0&&\psi_{1N}&
\dfrac{1}{2}\psi_{1N}&\dfrac{1}{2}\lambda_N\psi_{1N}
&\dfrac{1}{2}\lambda_N^2\psi_{1N}
\\
\psi_{21}& &0&-\frac 12 \psi_{21}&-\frac12\lambda_1 \psi_{21}&
-\frac 12 \lambda ^2_1\psi_{21}
\\
 &\ddots& & \vdots &\vdots &\vdots
 \\
0& &\psi_{2N}
&-\frac 12 \psi_{2N}&-\frac12\lambda_N \psi_{2N}&
-\frac 12 \lambda ^2_N\psi_{2N}
\\
0&\cdots&0 &0&0&\phi_{N+3}
\\
0&\cdots&0 &0&\phi_{N+3}&0
\\
0&\cdots&0& \phi_{N+3}&0&0
\end{array}\right.
\vspace{2mm} \\
\left.\begin{array}{ccc}
\qquad \dfrac{1}{2}\lambda_1^{3}\psi_{11}&\cdots
&\dfrac{1}{2}\lambda_1^{N+2}\psi_{11}\\
\qquad \vdots& &\vdots\\
\dfrac{1}{2}
\lambda_N^{3}\psi_{1N}
&\cdots&\dfrac{1}{2}
\lambda_N^{N+2}\psi_{1N}\\
\qquad -\dfrac{1}{2}\lambda_1^3\psi_{21}-\dfrac 12 \phi_{N+3}\psi_{11}
&\cdots &-
\dfrac12 \lambda _1^{N+2}\psi_{21}-\dfrac12
\lambda _1 ^{N-1}\phi_{N+3}\psi_{11}\\
\qquad \vdots & &\vdots\\
\qquad -\dfrac{1}{2}\lambda_N^3\psi_{2N}-\dfrac 12 \phi_{N+3}\psi_{1N}
&\cdots &-
\dfrac12 \lambda _N^{N+2}\psi_{2N}-\dfrac12
\lambda _N ^{N-1}\phi_{N+3}\psi_{1N}\\
\qquad 0& \cdots &0\\
\qquad 0& \cdots &0\\
\qquad 0& \cdots &0
\end{array}\right)
\vspace{2mm}\\
=(-1)^{N+1}\phi_{N+3}^3\prod ^N_{j=1}\psi_{1j}(-\lambda_j^3\psi_{2j}-
\frac12 \phi_{N+3}\psi_{1j})
\left|\begin{array}{cccc}
1&\lambda_1&\cdots&\lambda_1^{N-1}\\
\vdots&\vdots&&\vdots\\
1&\lambda_N&\cdots&\lambda_N^{N-1}
\end{array}\right|\not\equiv 0.\end{array}\right.$$

The above analysis allows us to conclude 
that the Hamiltonian systems (2.4), (2.9), (2.16) and
(2.19) are completely integrable in the Liouville 
sense$^{\cite{Arnold-BookMMCM1980}}$.

\section{Conclusion and remarks}

We have successfully extended binary nonlinearization of
AKNS spectral problem to higher-order symmetry constraints.  
For the first four orders $n=1,2,3,4$ of the symmetry constraints, 
we explicitly presented the Hamiltonian structures, the Lax representations,
$r$-matrices and 
independent integrals of motion in involution.
Binary nonlinearization under  
the symmetry constraint in the case of $n=3$ has also been considered
by Xu$^{\cite{Xu-JSMI1997}}$ but no Lax representations and $r$-matrix
has been presented till now.

Indeed, for the case of $n=2k+1$,
binary nonlinearization of AKNS spectral problem
can always be successfully handled.
However when $n$ is an even integer greater than $4$, we do not yet know
any concrete expression for new dependent variables which put
the constrained systems to be Hamiltonian.
We hope that we can find a rule to guide us to 
introduce new dependent variables such that the constrained systems
can be transformed into Hamiltonian forms. 

Throughout the paper, we suppose that the potentials $v,w$ 
and their derivatives with respect to $x$ tend to zero 
when ${|x|}\rightarrow\infty$. In the case 
without this zero boundary condition, the same problem
is very interesting to be considered. Binary
nonlinearization of  
$3\times 3$ matrix spectral problem$^{\cite{MaFO-PA1996}}$ or even
$n\times n$ matrix spectral problem$^{\cite{Xu-preprint1997}}$ 
for AKNS hierarchy
under higher-order symmetry constraints is another problem 
which needs to be investigated. We will deliver some analysis to 
the problems in a separate paper.

\section*{Acknowledgments}
One of the authors (Y.~S. Li) is grateful to 
City University of Hong Kong for kind invitation and warm 
hospitality. This work
is supported by 
the City University of Hong Kong,
the National Basic Research Project of Nonlinear Science, and the Ministry
of Education of China.

\end{document}